\documentclass[12pt,aps,prd,preprint,tightenlines,showpacs,nofootinbib,%
superscriptaddress]{revtex4-1}
\newcommand{\PRE}[1]{{#1}} 

\usepackage{hyperref}      
\usepackage{bm} 
\usepackage{amsmath} 
\usepackage{epsfig}
\usepackage{subfigure}
\usepackage{dsfont}

\begin{document}

\title{ \PRE{\vspace*{0.5in}}
Updated templates for the interpretation of LHC results on supersymmetry
in the context of mSUGRA \PRE{\vspace*{0.3in}} 
}

\author{K. Matchev, R. Remington}
\affiliation{Physics Department, University of Florida, Gainesville, Florida, U.S.A.
\PRE{\vspace*{.2in}}
}

\date{February 29, 2012}
\PRE{\vspace*{0.6in}}

\begin{abstract}
\PRE{\vspace*{.3in}} 
In this short note, we describe the preparation of updated templates for 
the interpretation of SUSY results from the LHC in the context of mSUGRA.  
The standard $(m_0,m_{1/2})$ plane is shown for fixed $\mu > 0$ and 
$m_{t}$ = 173.2 GeV. Two scenarios are considered:
(1) $A_0 = 0$ GeV and $\tan(\beta)=10$ and (2) $A_0 = -500$ GeV and 
$\tan(\beta)=40$. In each case, the universal scalar mass parameter 
$m_0$ varies in the range $m_0 \in \left[40,3000\right]$ GeV, 
while the universal gaugino mass parameter $m_{1/2}$ varies 
in the range $m_{1/2} \in \left[100,1000\right]$ GeV.
We delineate notable regions in parameter space, including
the region with a charged LSP (stau), the LEP2 reach, and the
cosmologically preferred region with 100\% neutralino dark matter. 
The templates also show mass contours 
for a few key particles (gluino, squark and Higgs boson).
The mass spectrum is calculated with the SoftSusy-3.2.4 package, 
while the neutralino relic density is obtained with MicrOMEGAs
version 2.4.
\end{abstract}


\maketitle

\section{Introduction}
\label{sec:introduction}

Low energy supersymmetry (SUSY) 
\cite{Golfand:1971iw,Ramond:1971gb,Neveu:1971rx,Neveu:1971iv,Volkov:1973ix,Wess:1973kz,Wess:1974tw,Fayet:1977yc}
is a primary target of the LHC collaborations in
their quest for new physics at the TeV scale. Unfortunately, even the most minimal supersymmetric
extension of the Standard Model comes with a large number of a priori unknown
parameters (superpartner masses, mixing angles and CP violating phases).
The experimental exploration of such a large parameter space is impractical.
Therefore, it has become customary to present experimental limits 
from searches for supersymmetry in terms of simple benchmark models with
very few input parameters, see e.g. 
\cite{Battaglia:2001zp,Allanach:2002nj,Battaglia:2003ab,Allanach:2006st,AbdusSalam:2011fc}.
The most popular SUSY benchmark model is the minimal supergravity (mSUGRA) model
\cite{Chamseddine:1982jx,Barbieri:1982eh,Hall:1983iz},
a.k.a. the ``constrained MSSM'' (cMSSM). In this model the SUSY mass parameters
are input at the grand unification (GUT) scale, conventionally defined as
the scale where the two gauge couplings $g_1$ and $g_2$ meet. At the GUT scale, all
scalars in the model have a common mass $m_0$, while all gaugino fermion 
superpartners have a common mass $m_{1/2}$. 
The model has two additional continuous parameters: $A_0$, a
common trilinear scalar coupling at the GUT scale, and $\tan(\beta)$, the ratio of the
two Higgs vacuum expectation values, plus a discrete parameter, the sign of the
higgsino mass parameter $\mu$. In order to obtain the physical SUSY mass spectrum,
the parameters $m_0$, $m_{1/2}$ and $A_0$ are evolved via the renormalization group equations (RGEs)
from the GUT scale down to the electroweak scale, where the radiatively corrected
mass spectrum is computed. Since the gauge and Yukawa couplings of the Standard Model
are input at the electroweak scale, the procedure must be iterated until it converges
on a stable solution. There are several state of the art, publicly available
codes on the market which can do these calculations, and they generally
give similar results in the bulk of the parameter space \cite{Allanach:2003jw}.
Here we use the SoftSusy-3.2.4 software package \cite{softsusy}.

\begin{table}[htb]
\begin{center}
\caption{Relevant input parameters for SoftSusy mass spectrum calculations}
\label{tab:params}
\begin{tabular}{ |c|c|c| }
\hline
Parameter &  \multicolumn{2}{c|}{Value}   \\ \hline  \hline
$\alpha_{s}(m_Z)$ & \multicolumn{2}{c|}{0.1184}   \\ 
$\alpha_{em}^{-1}(m_Z)$ & \multicolumn{2}{c|}{127.934}    \\
$m_{t}$  & \multicolumn{2}{c|}{173.2 GeV} \\
$m_{b}$  & \multicolumn{2}{c|}{4.2 GeV} \\ 
sign($\mu$) & \multicolumn{2}{c|}{+} \\ \hline
$A_0$ & 0 GeV & -500 GeV \\
$\tan(\beta)$ & 10 & 40 \\ \hline
\end{tabular}
\end{center}
\end{table}

In this note we provide details of the design of mSUGRA templates
which can be used for interpretation of supersymmetry searches 
at the LHC. Given the ubiquitousness of the mSUGRA model, it has 
become customary to present exclusion limits from SUSY searches
in the $m_0 - m_{1/2}$ parameter plane \cite{Baer:1994nc}, for fixed values of the 
remaining mSUGRA (and Standard Model) parameters (shown in Table~\ref{tab:params}).  
In anticipation of the improved reach with 2011 and 2012 data, 
we extend the previously considered ranges of $m_0$ and $m_{1/2}$,
allowing them to vary in $m_0 \in \left[40,3000\right]$ GeV and 
$m_{1/2} \in \left[100,1000\right]$ GeV, correspondingly.  
Two different $m_0 - m_{1/2}$ slices are considered. The first
has a moderate value of $\tan(\beta)=10$, while $A_0$
is taken to be $A_0=0$. The second $m_0 - m_{1/2}$ slice 
has a higher value of $\tan(\beta)=40$ and a different $A_0=-500$ GeV. 
In each case, the sign of the $\mu$ parameter is positive,
and the top mass is taken to be $m_t=173.2$ GeV.
In what follows, we refer to these two $m_0 - m_{1/2}$
scans by their respective $\tan(\beta)$ parameters.

\section{mSUGRA templates}

\begin{figure}[t!]
\begin{center}
\subfigure[~$\tan(\beta)=10$ (v1)]{\label{fig:tanb10_v1}\includegraphics[scale=0.27]{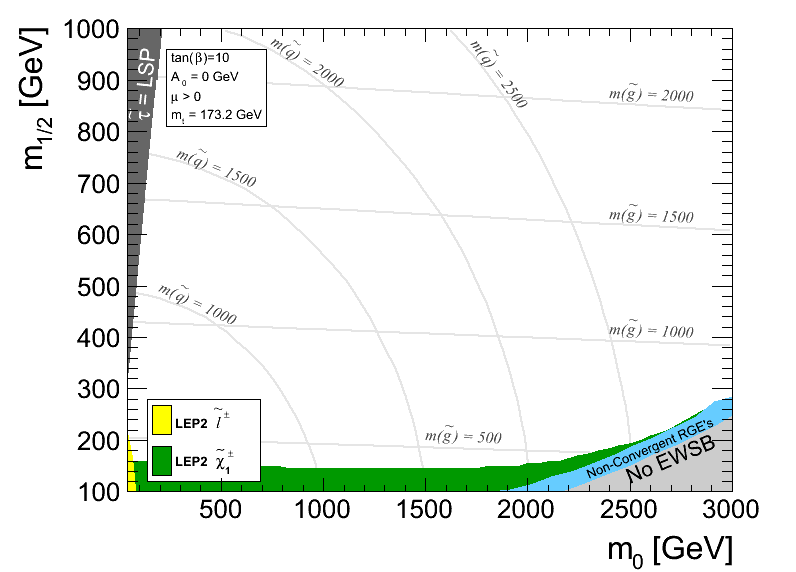}} 
\subfigure[~$\tan(\beta)=10$ (v2)]{\label{fig:tanb10_v2}\includegraphics[scale=0.27]{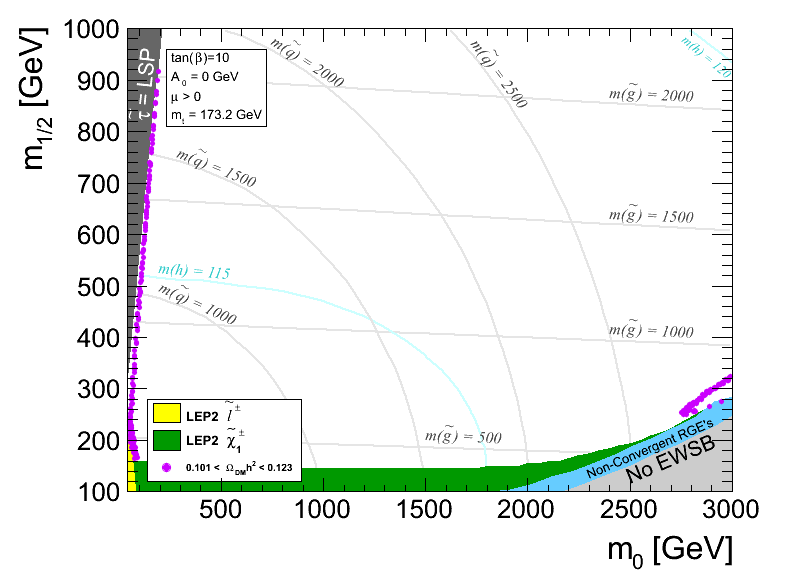}} \\
\caption{Template versions 1 (left) and 2 (right) for the $\tan(\beta) = 10$ scan. The features are described in the text. }
\label{fig:tanb10}
\end{center}
\end{figure}

\begin{figure}[t]
\begin{center}
\subfigure[~$\tan(\beta)=40$ (v1)]{\label{fig:tanb40_v1}\includegraphics[scale=0.27]{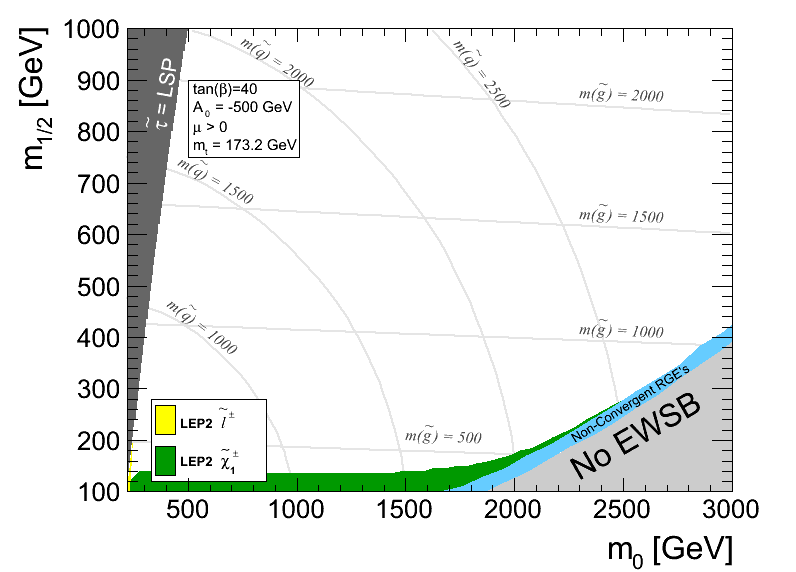}} 
\subfigure[~$\tan(\beta)=40$ (v2)]{\label{fig:tanb40_v2}\includegraphics[scale=0.27]{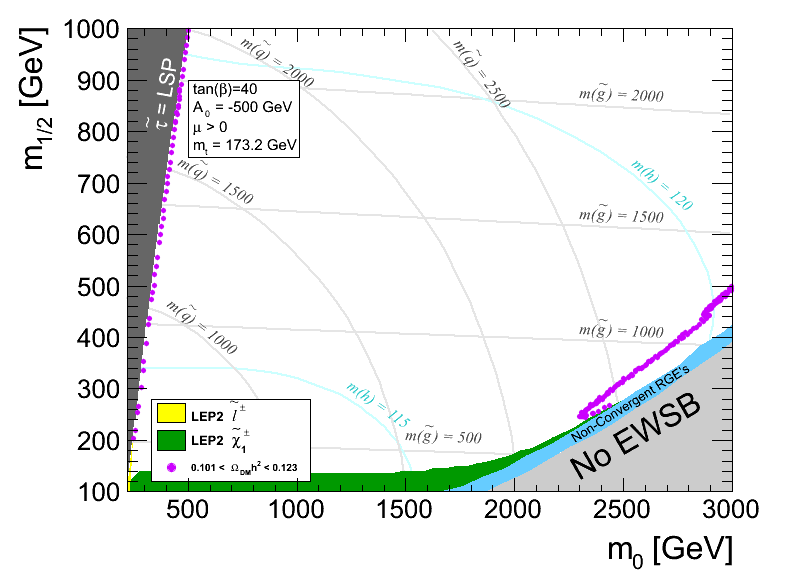}} \\
\caption{Template versions 1 (left) and 2 (right) for the $\tan(\beta) = 40$ scan. The features are described in the text. }
\label{fig:tanb40}
\end{center}
\end{figure}

We begin by presenting the templates in their final form 
before discussing the details of their respective features. 
The templates are displayed in Figs.~\ref{fig:tanb10} 
and \ref{fig:tanb40} for $\tan(\beta) = 10$ and $\tan(\beta) = 40$, respectively. 
The calculations are performed across an $m_0 - m_{1/2}$ grid. 
The LHC mSUGRA scans \cite{CMSscan} sample this grid in 20 GeV intervals 
for both $m_0$ and $m_{1/2}$ with about 10k events for each point. 
A finer granularity is of course desirable but is computationally 
prohibitive, as each grid point has to be propagated 
through the generation, simulation, and reconstruction chain. The 
storage space required for hosting these datasets would 
also grow quadratically with finer granularity. 
Since our goal is to simply provide a canvas which delineates 
theoretical and experimental regions of interest in the $m_0 - m_{1/2}$ plane, 
we only need to make use of the mass spectrum calculations, 
and these come in concise SLHA files from the SoftSusy output.  
This allows us to pursue a much finer granularity of 5 GeV intervals.
In the high $\tan(\beta)$ scan, grid points with $m_0 < 220$ GeV 
do not yield a suitable neutral dark matter candidate,
thus the scan effectively begins at $m_0=220$ GeV.

\section{Constraints on the mSUGRA parameter space}

Within the mSUGRA parameter space considered here, there are regions 
which are not considered to be viable for hypothesis testing, since they
are subject to constraints imposed by theory or experiment.

\subsection{Theoretical constraints}

\subsubsection{Charged LSP}

One nice feature of the mSUGRA model is that throughout most of the 
$m_0-m_{1/2}$ parameter space, the lightest supersymmetric 
particle (LSP) is a neutralino, which represents a potential dark matter candidate.
If a suitable discrete symmetry (such as $R$-parity) is imposed on the model,
the neutralino does become a dark matter candidate and its relic density can be
readily computed from the thermal freeze-out (see Sec.~\ref{sec:DM} below).
More importantly for collider searches though is the fact that 
neutralinos are very weakly interacting particles, and do not 
interact inside the detector. Thus mSUGRA parameter space points with neutralino 
LSP will be relevant for SUSY searches utilizing the missing transverse energy in the event.

As can be seen from Figs.~\ref{fig:tanb10} and \ref{fig:tanb40}, though,
at low values of $m_0$ there is a small corner in the $m_0-m_{1/2}$ plane 
(colored dark gray and denoted ``$\tilde\tau=LSP$''), where the lightest 
particle in the SUSY spectrum is a $\tau$ slepton (stau $\tilde\tau$). 
Being an electrically charged particle, the stau
is not a good dark matter candidate, therefore the dark gray regions 
in Figs.~\ref{fig:tanb10} and \ref{fig:tanb40} are 
cosmologically disfavored\footnote{This conclusion can be avoided 
if a) the stau is not the true LSP in the model, but slowly decays to a superWIMP such as the gravitino
\cite{Feng:2003xh} or b) if $R$-parity is violated and the stau decays to SM particles, with no
sizeable missing transverse energy in the event. In either case, the corresponding 
collider signatures fall into the scope of the exotica group.}.

\subsubsection{Correct electroweak symmetry breaking}

Another important theoretical requirement on the model is that the
global minimum of the scalar potential should break the electroweak symmetry, 
leaving QED and QCD intact. Within the $m_0-m_{1/2}$ plane one does find regions with the
wrong gauge symmetry breaking pattern. For example, at very large values of $m_0$,
one finds an area with no electroweak symmetry breaking, as indicated by an 
unphysical value of $\mu^{2} < 0$. In Figs.~\ref{fig:tanb10} and \ref{fig:tanb40}
that region is colored light gray and denoted ``No EWSB''.
(At very small values of $m_0$ at large $\tan(\beta)$, one may encounter an area with 
a charge-breaking vacuum, as signalled by a negative stau mass squared 
($M_{\tilde\tau}^2<0$) \cite{Feng:2005ba}. For simplicity, we do not specifically delineate 
that region, since it is already well inside the stau LSP region.)

\subsubsection{Convergence on reliable solutions to the RGEs}

As already mentioned in the Introduction, the numerical programs used to
solve the MSSM RGEs
use an iterative procedure, since some inputs
(the gauge and Yukawa couplings and $\tan(\beta)$) 
are specified at the weak scale, while others 
($m_0$, $m_{1/2}$ and $A_0$) are defined at the unification scale.
Given that the RGE's are very nonlinear functions,
one might expect that in some situations the iterative procedure 
will not converge and instead will exhibit chaotic behavior.
Indeed, this turns out to be the case in a narrow strip
(colored light blue), which is adjacent to the ``No EWSB'' region
in Figs.~\ref{fig:tanb10} and \ref{fig:tanb40}.

\begin{figure}[t]
\begin{center}
\subfigure[$~\tan(\beta)=10$]{\label{fig:NC10}\includegraphics[scale=0.3]{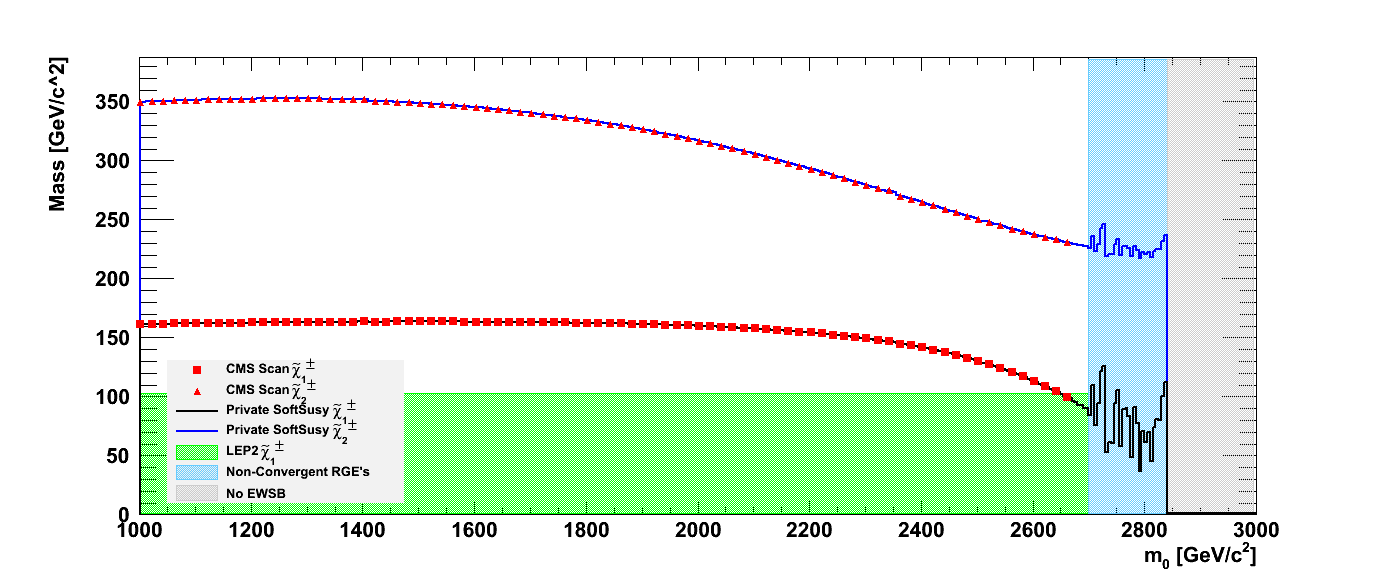}} \\
\subfigure[$~\tan(\beta)=40$]{\label{fig:NC40}\includegraphics[scale=0.3]{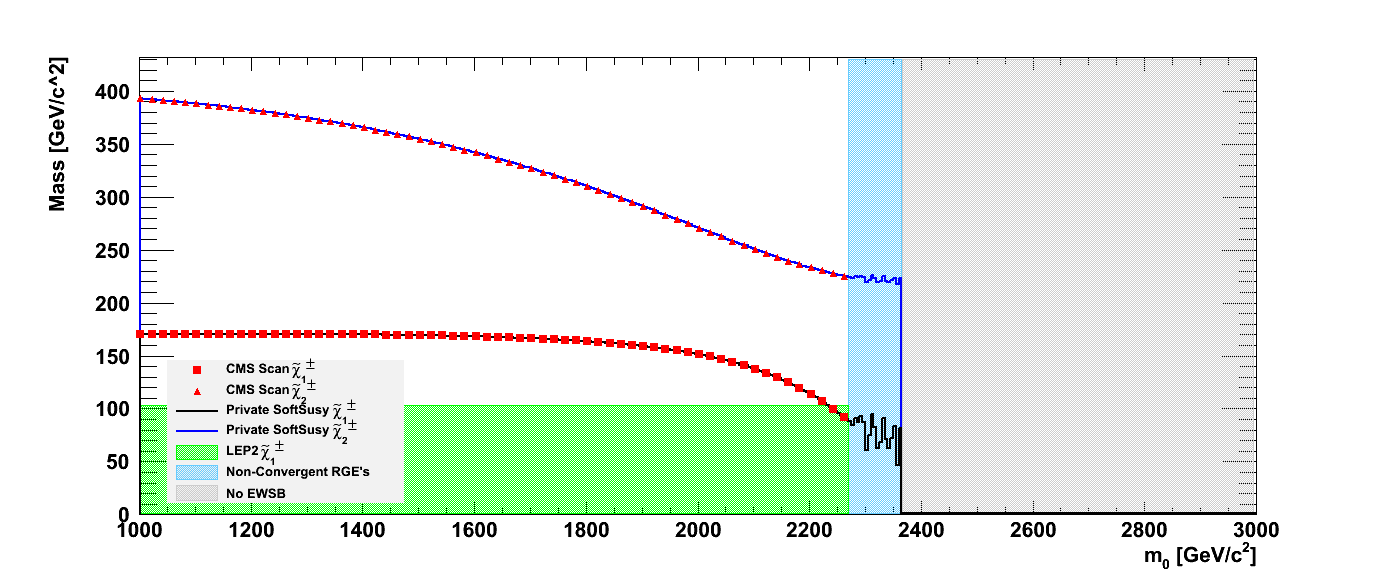}}
\caption{Masses of the two charginos $\tilde{\chi}^{\pm}_1$ and $\tilde{\chi}^{\pm}_2$ (in GeV)
as a function of $m_0$ for $m_{1/2} = 220$ GeV
 for the $\tan(\beta) = 10$ (top) and $\tan(\beta) = 40$ (bottom) scans. The non-convergent region 
is shaded in light blue, while the ``No EWSB'' region is shaded in gray. The green shaded 
region is ruled out by chargino searches at LEP.}
\label{fig:NC}
\end{center}
\end{figure}

Like all other SUSY spectrum calculators, SoftSusy checks for convergence 
after each iteration, and eventually terminates after a given fixed number 
of iterations\footnote{We have verified that increasing the maximum number 
of iterations does not alleviate the convergence problem, i.e. the chaotic 
behavior is intrinsically present in the system.}. 
One should keep in mind that even though SoftSusy warns the user of 
the non-convergence problem, it still provides a mass spectrum, 
albeit an unreliable one.
This is illustrated in Figs.~\ref{fig:NC10} and \ref{fig:NC40} 
which show the masses of the two charginos obtained in SoftSusy
as a function of $m_0$ for 
a horizontal slice across the $m_0-m_{1/2}$ planes of 
Figs.~\ref{fig:tanb10} and \ref{fig:tanb40} at a
fixed value of $m_{1/2} = 220$ GeV. Here, the region of 
RGE non-convergence as reported by SoftSusy is shaded in light blue. 
In this region, the expected value of $\mu$ as calculated from
minimizing the radiatively corrected effective potential,
is rather small, and the lightest chargino $\tilde\chi^+_1$
is higgsino-like. The oscillatory behavior of its mass $M_{\tilde\chi^+_1}$
at high $m_0$ is an indication of the convergence problem. 
Since the value of $\mu$ in the non-convergence region is small,
this region is always found to buffer 
the gray ``No EWSB'' region in which $\mu^{2} < 0$.  It is important 
to emphasize that non-convergence of the RGEs by itself does not 
necessarily mean that such points are ruled out or 
disfavored\footnote{Non-convergence of the RGEs means simply that --- the 
mass spectrum calculation is unreliable. In those cases, the 
SoftSusy manual \cite{softsusy} recommends using an extrapolation from the
neighboring regions where the program did converge.} --- they
may very well represent physically viable models and one should
not mistakenly incorporate them into the ``No EWSB'' region where 
$\mu^{2}$ becomes negative. Those are indeed two separate regions 
and deserve to be distinguished on the canvas.

In both Figs.~\ref{fig:NC10} and \ref{fig:NC40} we observe perfect agreement between 
the chargino masses as calculated by our private SoftSusy scan 
(solid lines, 5 GeV granularity) and those calculated by the
LHC scan \cite{CMSscan} (red markers, 20 GeV granularity). 
While for this slice of $m_{1/2} = 220$ GeV, the LHC scans 
do not extend into the regions where non-convergent RGE's are reported, 
it does happen in rare cases (a few out of a thousand) ---
the LHC scan would sample a point that falls just over the
boundary of the region of non-convergence. We have checked that
the chargino masses obtained in those cases do coincide with
the result that one would obtain by extrapolating from
the convergent region, so no ill-effects are expected from this occasional trespassing.  
 
It should be noted that while all of these theoretical constraints are ascertained
with the SoftSusy spectrum calculator, they are generally understood and expected 
constraints and appear when using other spectrum  calculators as well 
(Suspect \cite{suspect}, SPheno \cite{spheno}, IsaJet \cite{isajet}, etc.).

\subsection{Experimental Constraints}

Results from previous experiments can also be interpreted
in the context of the mSUGRA parameter space shown in Figs.~\ref{fig:tanb10} and \ref{fig:tanb40}.

\subsubsection{Chargino mass limits from LEP2}

\begin{figure}[htp]
\begin{center}
\subfigure[~$\tan(\beta)=10$]{\label{fig:CharginoMass10}\includegraphics[scale=0.28]{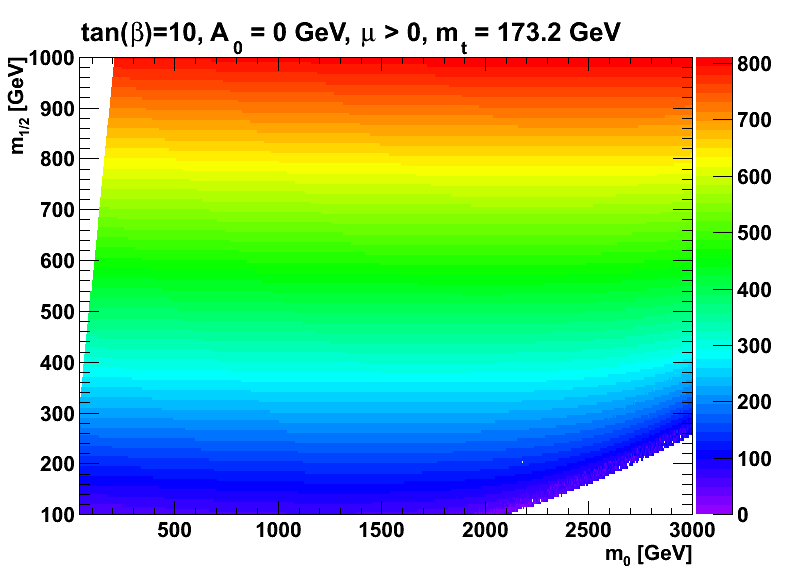}} 
\subfigure[~$\tan(\beta)=40$]{\label{fig:CharginoMass40}\includegraphics[scale=0.28]{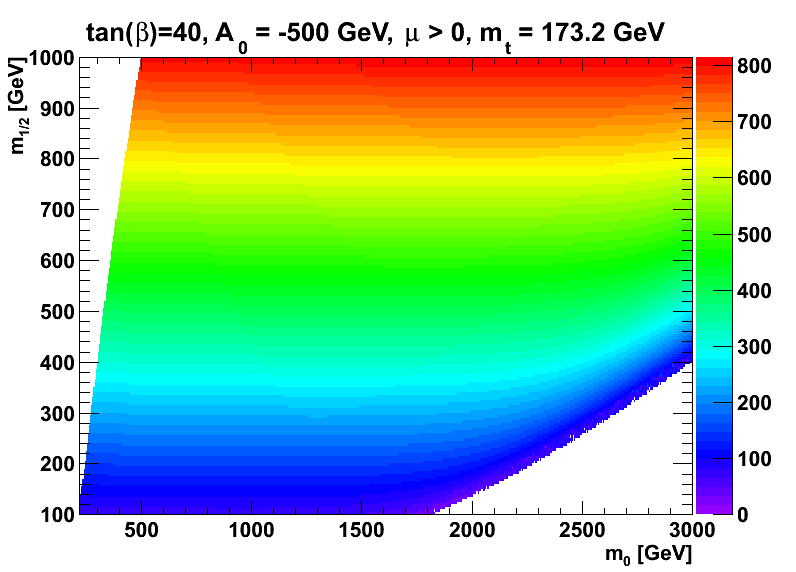}}\\
\caption{Lightest chargino mass $M_{\tilde\chi^+_1}$ 
as a function of $m_0$ and $m_{1/2}$ for (a) $\tan(\beta)=10$
and (b) $\tan(\beta)=40$.}
\label{fig:CharginoMass}
\end{center}
\end{figure}

LEP2 searches for chargino pair production are probing the two
areas of the $m_0-m_{1/2}$ plane which exhibit a light chargino
(at low $m_{1/2}$, where $\tilde\chi^+_1$ is wino-like, and 
at large $m_0$ where $\mu$ becomes small and $\tilde\chi^+_1$ is higgsino-like). 
In Figs.~\ref{fig:tanb10} and \ref{fig:tanb40} the corresponding 
regions are colored green and can be understood as excluded at 
95\%  C.L. by LEP2 \cite{PDG}. 
For reference, the lightest chargino mass $M_{\tilde\chi^+_1}$ 
is plotted across the $m_0-m_{1/2}$ grid in Figs.~\ref{fig:CharginoMass10} 
and \ref{fig:CharginoMass40} for $\tan(\beta) = 10$ and $\tan(\beta) = 40$,  
respectively.  

\subsubsection{Slepton mass limits from LEP2}

LEP2 has also performed searches for direct slepton production. 
Explicitly, the following exclusion limits are imposed \cite{PDG}:
\begin{eqnarray}
m(\tilde{e}_{R}) < 100 \mbox{ GeV } &\land& m(\tilde{\chi}^{0}_{1}) < 85 \mbox{ GeV } \\
m(\tilde{\mu}_{R}) < 95 \mbox{ GeV } &\land& m(\tilde{\mu}_{R}) - m(\tilde{\chi}^{0}_{1}) > 5 \mbox{ GeV } \\
m(\tilde{\tau}_{1}) < 86 \mbox{ GeV } &\land& m(\tilde{\tau}_{1}) - m(\tilde{\chi}^{0}_{1}) > 7 \mbox{ GeV } 
\end{eqnarray}  
The relatively small region which satisfies this condition can be found at
low $m_0$ and low $m_{1/2}$ and is colored yellow 
in Figs.~\ref{fig:tanb10} and \ref{fig:tanb40}.
It can be understood as excluded at 95\% C.L. by LEP2. 

\subsubsection{Dark matter relic density}
\label{sec:DM}

\begin{figure}[htp]
\begin{center}
\subfigure[$~\tan(\beta)=10$]{\label{fig:Omega10}\includegraphics[scale=0.28]{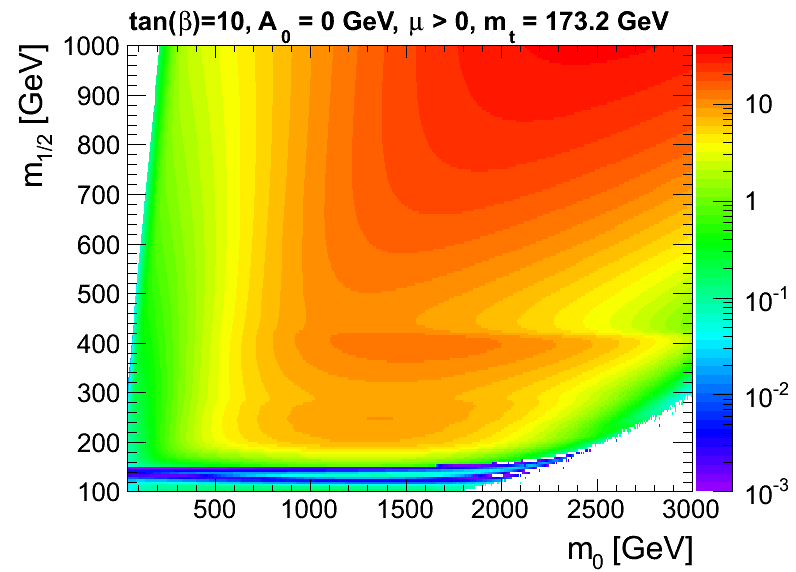}} 
\subfigure[$~\tan(\beta)=40$]{\label{fig:Omega40}\includegraphics[scale=0.28]{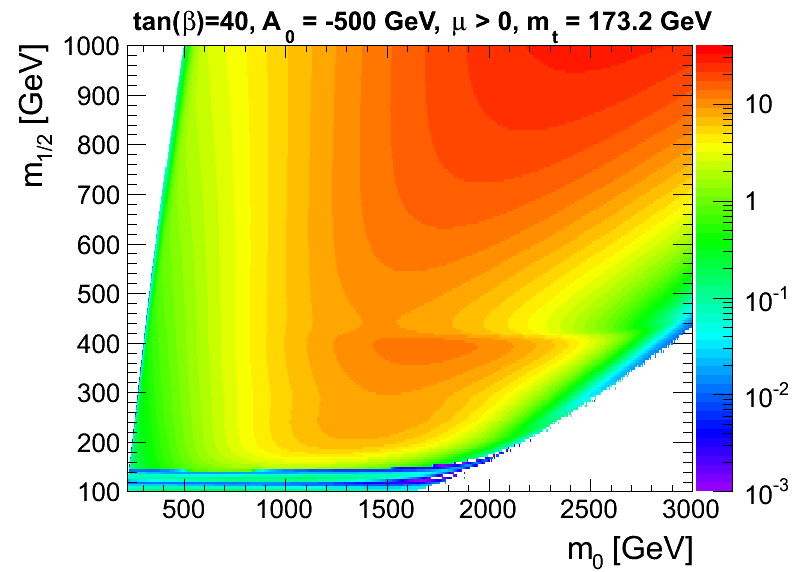}}\\
\caption{Dark matter relic density $\Omega h^{2}$ as a function of $m_0$ and $m_{1/2}$ for $\tan(\beta)=10$
 (top) and $\tan(\beta)=40$ (bottom).}
\label{fig:Omega}
\end{center}
\end{figure}

In addition to collider constraints from direct searches for superpartners, one may 
also wish to consider the cosmological constraint on the dark matter relic density
(which is indirect and perhaps more speculative, since it involves certain assumptions 
about the composition of the dark matter as well as the thermal history of the Universe).
We calculate the neutralino relic density across the $m_0-m_{1/2}$ grid 
with MicrOMEGAs version 2.4 \cite{Belanger:2010gh} using SoftSusy-3.2.4 for mass spectrum calculations.  
The resulting $\Omega_{\chi} h^{2}$ is displayed in Figs.~\ref{fig:Omega10} 
and \ref{fig:Omega40} for $\tan(\beta) = 10$ and $\tan(\beta) = 40$,  respectively. 
One can see that throughout most of the parameter space, the neutralino relic density 
is quite larger than the WMAP preferred range \cite{Komatsu:2010fb}
\begin{eqnarray}
0.102 < \Omega_\chi h^{2} < 0.123. 
\label{eq:DM}
\end{eqnarray}
Model points satisfying eq.~(\ref{eq:DM}) yield the appropriate amount of dark matter 
to within 3$\sigma$ of the measured value of 0.1123 $\pm$ 0.0035
and are denoted by magenta filled circles in version 2 of the templates
on Figs.~\ref{fig:tanb10}(b) and \ref{fig:tanb40}(b).
As expected, the proper relic density is obtained in 
two areas: 1) the neutralino-stau coannihilation region \cite{Ellis:1998kh} 
near the boundary of the ``$\tilde\tau$ LSP'' region 
and 2) the ``focus-point'' region at large $m_0$ 
\cite{Feng:1999mn,Feng:1999zg,Chan:1997bi}, where 
the neutralino LSP picks up a non-negligible higgsino component
\cite{Feng:2000gh}.
Other notable features on the plot in Fig.~\ref{fig:Omega}(a)
include the opening of the neutralino annihilation channel into
pairs of top quarks (near $m_{1/2}=430$ GeV)
and pairs of Higgs and/or gauge bosons (in the neighborhood of $m_{1/2}=300$ GeV).

\section{Mass spectrum}

\begin{figure}[htp]
\begin{center}
\subfigure[$~\tan(\beta)=10$]{\label{fig:GluinoMass10}\includegraphics[scale=0.28]{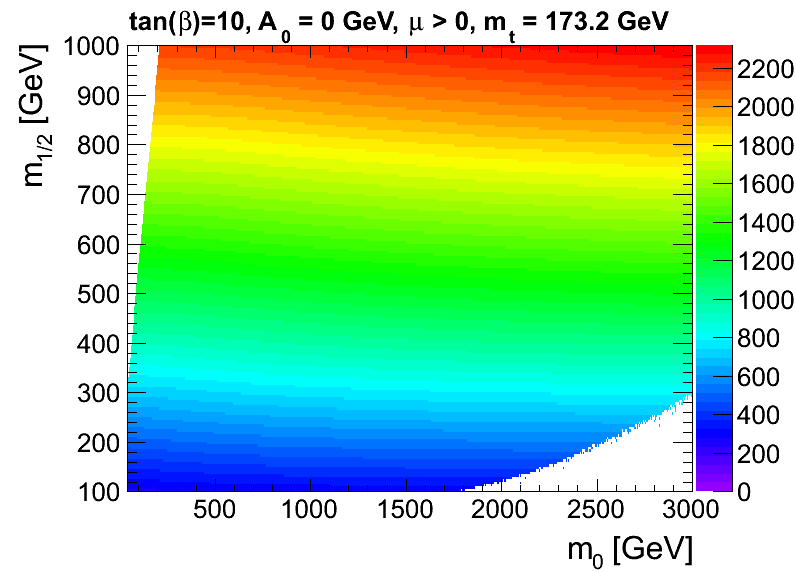}} 
\subfigure[$~\tan(\beta)=40$]{\label{fig:GluinoMass40}\includegraphics[scale=0.28]{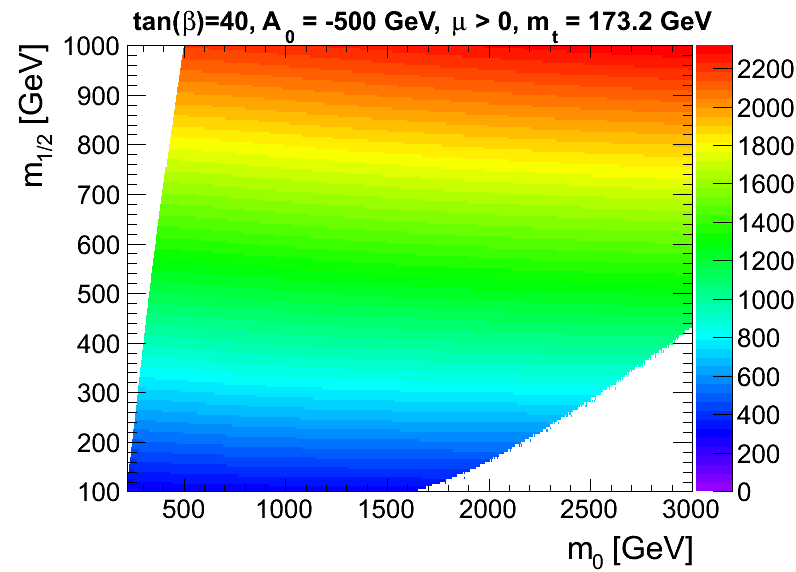}}\\
\caption{Gluino mass in GeV as a function of $m_0$ and $m_{1/2}$ for (a) $\tan(\beta)=10$ 
and (b) $\tan(\beta)=40$.}
\label{fig:GluinoMass}
\end{center}
\end{figure}

\begin{figure}[htp]
\begin{center}
\subfigure[$~\tan(\beta)=10$]{\label{fig:SquarkMass10}\includegraphics[scale=0.28]{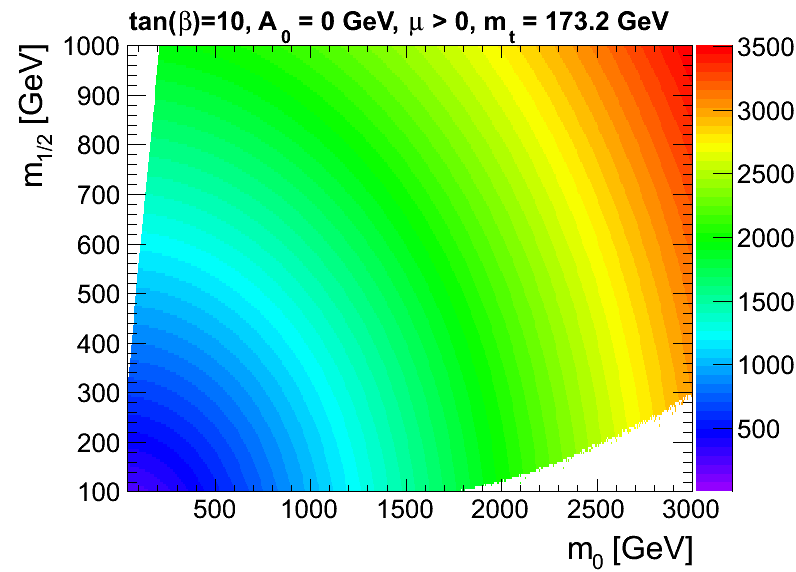}} 
\subfigure[$~\tan(\beta)=40$]{\label{fig:SquarkMass40}\includegraphics[scale=0.28]{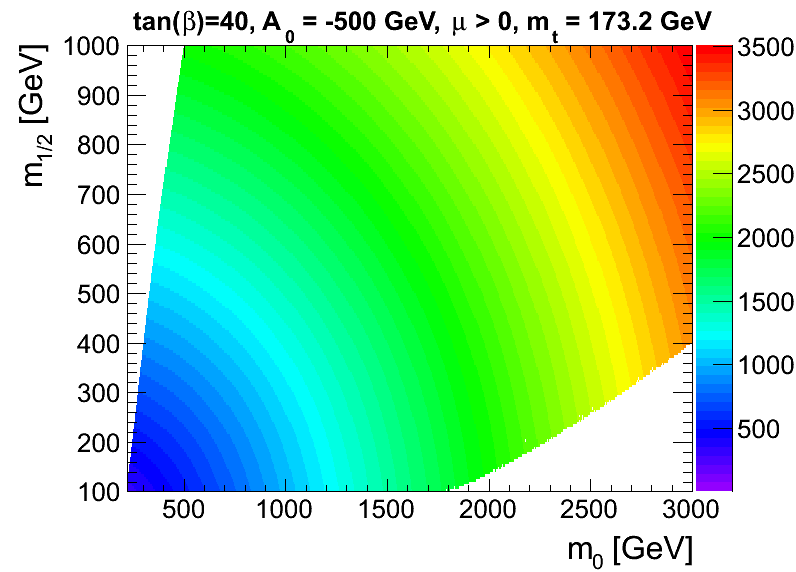}} \\
\caption{Average first generation squark mass in GeV as a function of $m_0$ and $m_{1/2}$ for 
(a) $\tan(\beta)=10$ and (b) $\tan(\beta)=40$.}
\label{fig:SquarkMass}
\end{center}
\end{figure}

\begin{figure}[htp]
\begin{center}
\subfigure[$~\tan(\beta)=10$]{\label{fig:HiggsMass10}\includegraphics[scale=0.28]{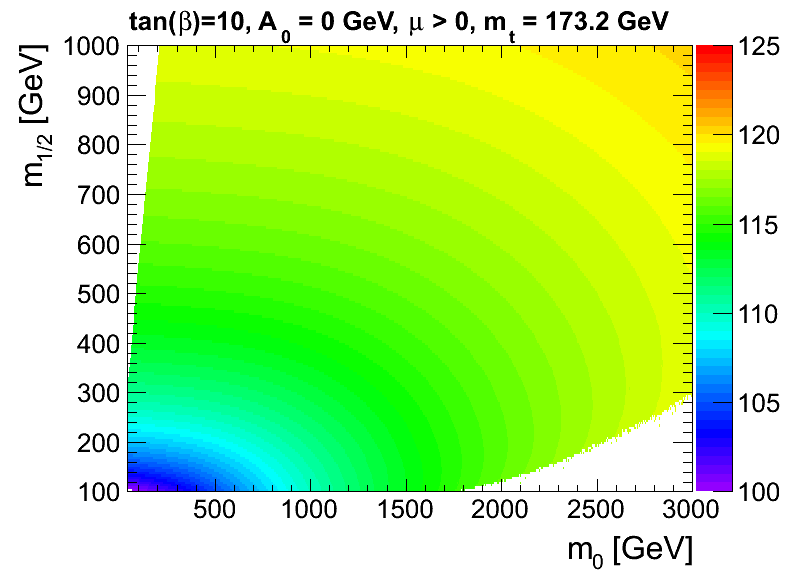}} 
\subfigure[$~\tan(\beta)=40$]{\label{fig:HiggsMass40}\includegraphics[scale=0.28]{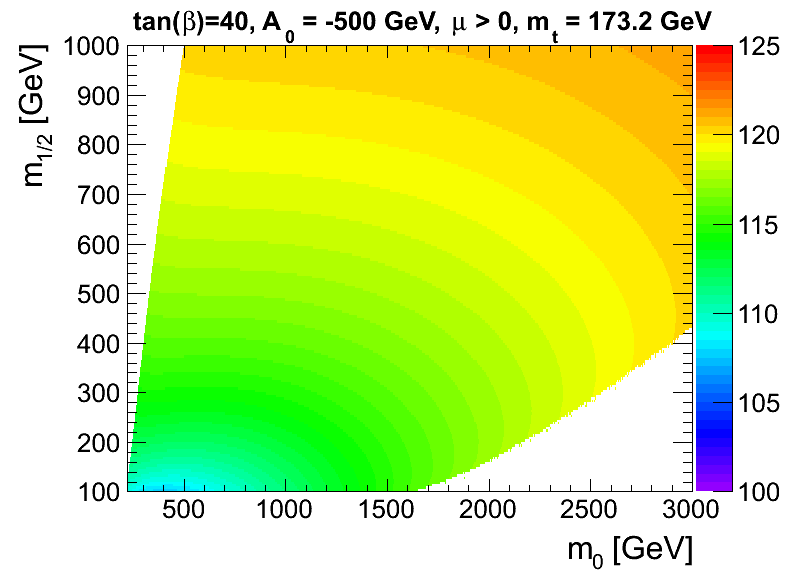}} \\
\caption{Higgs mass in GeV as a function of $m_0$ and $m_{1/2}$ for (a) $\tan(\beta)=10$ and (b) $\tan(\beta)=40$.}
\label{fig:HiggsMass}
\end{center}
\end{figure}

It is a tradition to display a few meaningful isomass contours for 
particles of interest on the mSUGRA canvas. 
 In the past, the gluino and squark masses were displayed in 250 GeV 
intervals.\footnote{Here we take the average of the four first generation squarks.  
These are observed to always be within 8\% of one another.}  
We continue to display these contours, opting for 500 GeV 
intervals to reduce congestion. In Version (2) of the templates we go further
and provide two additional contours for a Higgs mass of 115 and 120 GeV, 
respectively. These values are particularly relevant at this stage as 
the results of the Higgs searches at the LHC may 
soon make contact with this parameter space \cite{Feng:2011aa}.  
As a reference, Figs.~\ref{fig:GluinoMass}, 
\ref{fig:SquarkMass}, and \ref{fig:HiggsMass} show the masses of the gluino, squark (average), 
and light CP-even Higgs boson, respectively, across the $m_0-m_{1/2}$ plane.

\section{Application}

It is expected that Version (1) of the templates is used for public presentation.  
Both versions, however, are available as TCanvas objects and C-macros in the ROOT framework.  
These can be downloaded from the web at
\begin{itemize}
\item {\bf $\tan(\beta) = 10$}\\ 
 tier2.ihepa.ufl.edu/\textasciitilde remington/SUSY/mSUGRA/GridTanb10\_v1.root (.C)\\
 tier2.ihepa.ufl.edu/\textasciitilde remington/SUSY/mSUGRA/GridTanb10\_v2.root (.C)
\item {\bf $\tan(\beta) = 40$}\\
 tier2.ihepa.ufl.edu/\textasciitilde remington/SUSY/mSUGRA/GridTanb40\_v1.root (.C)\\
 tier2.ihepa.ufl.edu/\textasciitilde remington/SUSY/mSUGRA/GridTanb40\_v2.root (.C)
\end{itemize}
and are also available as ancillary files from the arXiv version of this note.

The following code can be used in ROOT (v.27) to draw the canvas from the root files above:

\begin{center}
\begin{verbatim}
TFile f(``GridTanb10_v1.root'');
TCanvas *c = (TCanvas*) f.Get(``GridCanvas'');
c->Draw();
MyExclusionContour->Draw(``SAME'');
\end{verbatim}
\end{center}

\begin{figure}[htp]
\begin{center}
\includegraphics[scale=0.39]{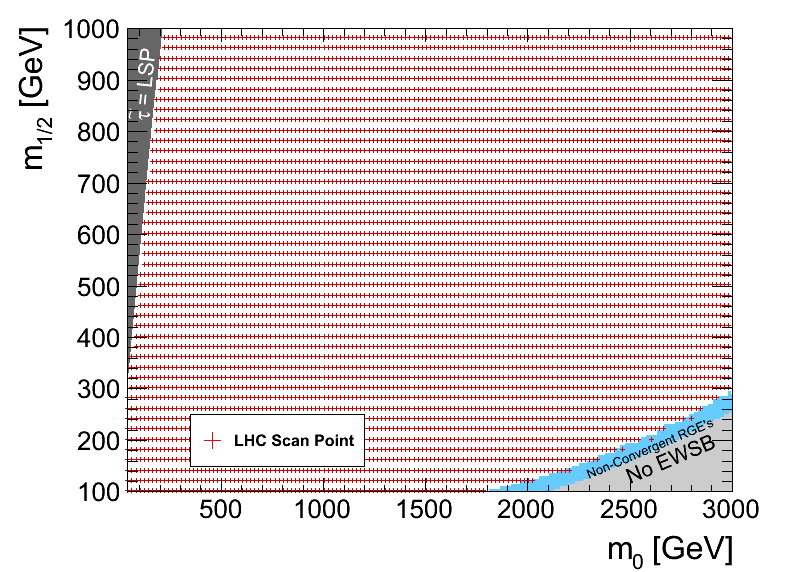} \\
\caption{LHC scan points (+) overlaying the mSUGRA template for $\tan(\beta) = 10$.}
\label{fig:tanb10_ScanPoints}
\end{center}
\end{figure}

\begin{figure}[htp]
\begin{center}
\includegraphics[scale=0.39]{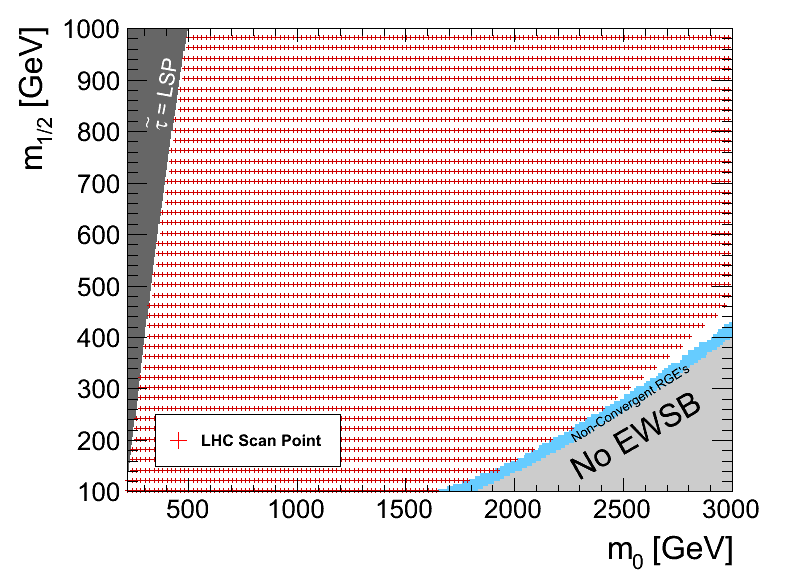} \\
\caption{LHC scan points (+) overlaying the mSUGRA template for $\tan(\beta) = 40$.}
\label{fig:tanb40_ScanPoints}
\end{center}
\end{figure}

As a final check, we survey the model points that are expected to be in the official LHC scans
\cite{CMSscan} and we represent them symbolically with a ``+'' on 
top of our templates to assess the coverage.  
This can be seen in Figs.~\ref{fig:tanb10_ScanPoints} and \ref{fig:tanb40_ScanPoints} 
for $\tan(\beta) = 10$ and $\tan(\beta) = 40$, respectively. 
For $\tan(\beta) = 10$ we see that there is full coverage everywhere. 
For $\tan(\beta) = 40$, however, we see a small envelope in between 
the dark matter preferred region and the non-convergent RGE region
where the LHC scan does not sample. Our recommendation is that future 
scans should be extended to include this interesting region.
Analysts should be prepared to perform a small extrapolation in this 
region in the case that their exclusion contour terminates nearby.

\section*{Acknowledgments}
We thank the CMS SUSY group conveners for encouragement
and S. Mrenna and S. Padhi for useful discussions.
We also thank J.~L.~Feng for pointing out some inconsistencies in
the previous mSUGRA templates used by CMS and ATLAS.

\bibliography{AN-12-005}{}

\end{document}